# ELF: An End-to-end Local and Global Multimodal Fusion Framework for Glaucoma Grading

Wenyun Li
*Department of Computer and Information Science*
*University of Macau*
Macau, China
mc05411@um.edu.mo

Chi-Man Pun[*]
*Department of Computer and Information Science*
*University of Macau*
Macau, China
cmpun@umac.mo

*Abstract*—Glaucoma is a chronic neurodegenerative condition that can lead to blindness. Early detection and curing are very important in stopping the disease from getting worse for glaucoma patients. The 2D fundus images and optical coherence tomography(OCT) are useful for ophthalmologists in diagnosing glaucoma. There are many methods based on the fundus images or 3D OCT volumes; however, the mining for multi-modality, including both fundus images and data, is less studied. In this work, we propose an end-to-end local and global multi-modal fusion framework for glaucoma grading, named ELF for short. ELF can fully utilize the complementary information between fundus and OCT. In addition, unlike previous methods that concatenate the multi-modal features together, which lack exploring the mutual information between different modalities, ELF can take advantage of local-wise and global-wise mutual information. The extensive experiment conducted on the multi-modal glaucoma grading GAMMA dataset can prove the effiectness of ELF when compared with other state-of-the-art methods.

## I. INTRODUCTION

Glaucoma is one of the common causes of irreversible but preventable blindness in senior populations [1]. As the second-leading cause of blindness after cataracts, glaucoma has affected more than 70 million patients worldwide. Furthermore, many kinds of glaucoma have no precursor. Prompt detection and treatment of glaucoma can considerably prevent blindness. Ophthalmologists have mainly approached fundus photographs and optical coherence tomography (OCT) diagnosing glaucoma. Fundus photograph is cheap and intuitive since it can clearly show the optic disc, cup, and blood vessels. As a new 3D imaging technology, OCT can display 3D images of the human eyes. Many methods utilise fundus photographs [2] or OCT volumes [3], for diagnosing glaucoma. However, the method that employs the two modalities simultaneously is less studied.

Multi-modality learning is instrumental in the medical imaging field for the different modal data can enhence the complementary information [4]–[6]. In general, modal medical images can complement unimodal data in a multiview way. Nevertheless, modal heterogeneity is always an arduous task to solve. For cross-over the heterogeneous gap of multi-modality, there are many methods focusing on the fusion of medical imaging. Typically, the fusion of multimodal medical images can be classified as early fusion and later fusion [7]. Wang et al. [8] proposed a medical image fusion method based on the Laplacian pyramid decomposition and adaptive sparse representation to fuse the brain slice of sarcoma images. Li et al. [9] proposed a non-subsampled contourlet transform-based medical image fusion algorithm to fusion the CT and MRI images. Dinh et al. [10] utilizes the Equilibrium optimizer algorithm and local energy functions to fuse two CT images.

However, although these methods achieve remarkable performance on modal medical image fusion, there are still some issues to solve. They mainly process the multi-modality data in a single-modality way, which means that the mutual information between different modals is ignored. Moreover, the simple concatenating without thinking of the unbalanced distribution of multi-modal data will reduce the preciousness of the multi-modal task. Thus, to solve these problems, we propose An End-to-end Local and Global Multi-modal Fusion Framework for Glaucoma Grading(ELF).

In comparison with previous work, our main contributions can be summarized as follows:

1) We propose an end-to-end local and global multi-modal fusion framework for glaucoma grading(ELF), which can simultaneously take advantage of 2D fundus photographs and 3D OCT volumes. This multi-modal learning-based glaucoma grading can perform better than any single-modality grading method thanks to the complementary information between different eye data.
2) We employ the attention mechanism in the local-wise module and global-wise module to mine the mutual information between different modalities. These can make the features concentrate on the shared information to achieve better performance.
3) We have conducted extensive experiments on multi-modality GAMMA datasets. The results show that the proposed ELF is superior to the state-of-the-art methods in glaucoma grading.

[*]Corresponding author.

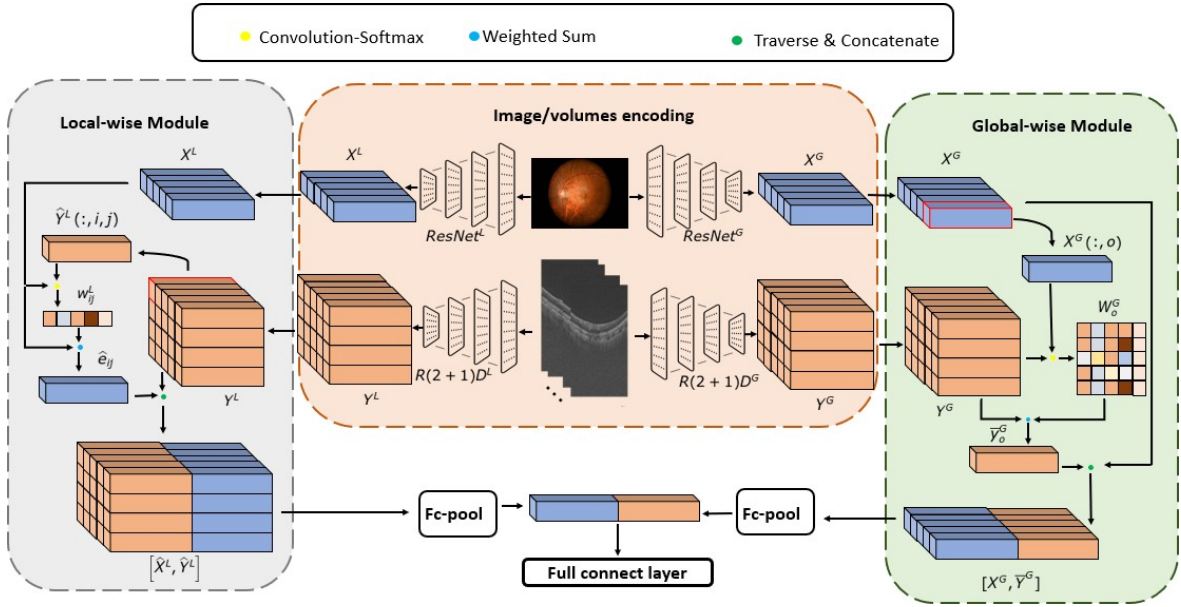

Fig. 1: The framework of the proposed End-to-end Local and Global Multi-modal Fusion Framework for Glaucoma Grading (ELF).

## II. METHODS

The previous works rarely explore the mutual information between 2D fundus images and the 3D OCT volumes. As a significant novelty, we mine the local-wise and the global-wise mutual relation between the two multimodal data to boost the discriminative ability. The overall framework of ELF is presented in Fig.1.

### A. Image/3D volumes encoding

Multi-modal medical images can describe the structure of glaucoma in a multiview way. Unlike previous works [11], [12] that regard the 3D OCT volumes as 2D images or transform the 3D OCT volumes into thickness map [13], which would lose the spatial structure information, we employ the pretrained R(2+1)D [14] model to extract the features from 3D volumes. Moreover, for 2D fundus images, we adopt the pretrained ResNet [15] as the backbone to learn the features. In order to construct the LM and the GM, we utilise two separate ResNets for image data $x$ as follows,

$$\mathbf{X}^L = ResNet^L(x) \qquad \mathbf{X}^G = ResNet^G(x) \qquad (1)$$

where $\{\mathbf{X}^L, \mathbf{X}^G\} \in \mathbb{R}^{C_x \times H_x \times W_x}$. Similar operation to 3D volumes data $y$ as follows,

$$\mathbf{Y}^L = R(2+1)D^L(y) \qquad \mathbf{Y}^G = R(2+1)D^G(y) \qquad (2)$$

where $\{\mathbf{Y}^L, \mathbf{Y}^G\} \in \mathbb{R}^{T_y \times C_y \times H_y \times W_y}$. For 2D and 3D data, we reshape the $\mathbf{X}$ and $\mathbf{Y}$ matrices to $\{\mathbf{X}^L, \mathbf{X}^G\} \in \mathbb{R}^{C_x \times H_x \times W_x}$ and $\mathbf{Y}^L, \mathbf{Y}^G \in \mathbb{R}^{T_y \times C_y \times H_y \times W_y}$.

### B. Local-wise Module

Unlike previous methods that separate the fundus images and scanning volumes, we argue that it is necessary to consider the 2D and 3D data jointly. Here we employ the attention mechanism to mine the relationship between multi-modal data. Specifically, in order to facilitate the attention calculation, we adopt the 1 × 1 convolution on the data $\mathbf{Y}^L$ as follows,

$$\mathbf{Y}^L = Conv(\mathbf{Y}^L) \qquad (3)$$

where $\mathbf{Y}^L \in \mathbb{R}^{C_x H_x \times C_y H_y \times W_y}$. It denotes that $\mathbf{Y}^L(:, i, j)$ is transformed into the same space with the 2D data $\mathbf{X}^L$. Then, to derive the attention weight distribution, we take the convolution operation with every data $\mathbf{Y}^L(:, i, j)$, we have

$$\mathbf{w}_{ij}^L = softmax\left(\frac{\mathbf{X}^L \otimes \mathbf{Y}^L(:, i, j)}{\tau^L}\right) \qquad (4)$$

where $\otimes$ denotes the convolution operation. $\tau^L$ is the temperature factor. With the help of a weight distribution vector $\mathbf{w}_{ij}^L \in \mathbb{R}^{W_x}$, we can get the weighted fundus images representation as follows,

$$\mathbf{e}_{ij} = \sum_{l=1}^{W_x} \mathbf{w}_{ij}^L(l) \cdot \mathbf{X}^L(:, l) \qquad (5)$$

To summarise, we have the fundus image representation as follows

$$\mathbf{\hat{X}}^L(:, i, j) = \mathbf{e}_{ij} \qquad (6)$$

where $\mathbf{\hat{X}}^L \in \mathbb{R}^{C_x H_x \times C_y H_y \times W_y}$. Then we concatenate the transformed 3D and 2D representations over the first dimension.

## C. Global-wise Module

GM follows a similar paradigm to the LM. Rather than LM focusing on the attention between the modified image and the 3D local feature maps, GM attend to concentrate on the relation of image representation and the OCT global features.

Similar to the LM, we first adopt a $1 \times 1$ convolution layer to transform the feature $\mathbf{Y}^G \in \mathbb{R}^{T_y \times C_y H_y \times W_y}$ to $\mathbf{Y}^G \in \mathbb{R}^{C_x H_x \times C_y H_y \times W_y}$ for attention weight calculation. Then we convolve the 3D representation $\mathbf{Y}^G$ with the image representation $\mathbf{X}^G(:,o)$ as follows,

$$\mathbf{W}_o^G = softmax\left(\frac{\mathbf{Y}^G \otimes \mathbf{X}^G(:,o)}{\tau^G}\right) \quad (7)$$

where $\tau^G$ is the temperature factor. The weight map $\mathbf{W}_o^G$ denotes the importance of the 3D representation towards the $o$-th image representation. Similar to LM, we get the weighted OCT volumes representation as follows,

$$\overline{\mathbf{y}}_o^G = \sum_{i=1}^{W_y} \sum_{j=1}^{C_y H_y} \mathbf{W}_o^G(i,j) \cdot \mathbf{Y}^G(:,i,j) \quad (8)$$

We have $\overline{\mathbf{Y}}^G = [\overline{\mathbf{y}}_1^G, \overline{\mathbf{y}}_2^G, \cdots, \overline{\mathbf{y}}_{C_x H_x}^G] \in \mathbb{R}^{C_x H_x \times W_x}$. Then we do the similar concatenate operation as LM.

## D. Multi-modality fusion

After the local-wise and global-wise learning, we generate two discriminative features with mutual information. Then we process two features with the classifiers to get the same dimensional features. Finally, we concatenate the features and feed them into a subsequent fully-connected layer.

## III. EXPERIMENTS

### A. Dataset

The GAMMA dataset was provided by [11], which contains 100 samples of fundus images and OCT volumes paired with three-level glaucoma grading("None", "Early" and "Mid-Advanced") labelled by experienced ophthalmologists. Each OCT volume contains $256 \times 992 \times 512$ pixels, of which 256 is a two-dimensional cross-sectional number. The fundus images are collected with a resolution of $2000 \times 2992$ pixels. The dataset concentrates on glaucoma grading via multi-modality data.

### B. Implementation Details

The experiment is conducted by PyTorch on NVIDIA TESLA V100 GPUs. Before the training, we first resized the fundus images to $1024 \times 1024$, then resized each layer of OCT volumes to $384 \times 384$. For the fundus image encoder, we select ResNet50 [15] as the backbone; the inter-mediate representations of fundus images have the shape of $2048 \times 14 \times 14$ in Eqn.1. For 3D volumes encoder, we select pretrained R(2+1)D [14] as the backbone to extract the spatial information including label information.

In our experiment, we set the batch size to 8 and employed Adam as our optimizer with an initial learning rate of 0.001.

TABLE I: Comparisons between ELF and other SOTA methods on glaucoma grading. The highest score is shown in **boldface**.

| Methods | Dataset | Acc | Kappa |
|---|---|---|---|
| Gabriel et al. [17] | private | 0.818 | 0.719 |
| Cheng et al. [18] | private [19] | 0.798 | 0.641 |
| Parashar et al. [20] | HRF [21] | 0.863 | 0.785 |
| Corolla [13] | GAMMA [11] | 0.900 | 0.855 |
| ELF | GAMMA [11] | **0.930** | **0.896** |

Temperature factors $\tau^L$ and $\tau^G$ in Eqn.4 and Eqn.7 are set to 6.0 and 4.0, respectively. For the fundus image data augmentation, we utilise a combination of random resized cropping, random horizontal flipping, random vertical flipping and random rotation to boost the robust for noise, for 3D OCT volumes data, we augment data with a combination of centre cropping, random horizontal flipping and normalization. We concatenate two 1024 features in the fusion stage to jointly grade glaucoma. We set our epoch number as 1000, and use the cross entropy loss as our loss.

### C. Evaluation

In our experiment, we utilise two evaluation matrics includes, accuracy (Acc) and Cohen's Kappa coefficient (Kappa) [16], respectively. For the glaucoma grading case, the higher Acc and Kappa correspond to better prediction performance. For the ordered categories in our experiment, quadratically weighted Kappa can manifest the different extents of the error. So we use Kappa to evaluate the effectiveness of the proposed method.

### D. Results

We compare our proposed ELF method with other state-of-the-art glaucoma grading methods. Gabriel et al. [17] evaluate their methods in a single-modality setting on their private datasets with more than 1000 samples. Cheng et al. [18] mainly focuses on glaucoma grading based on retcam images, they conducted their experiment on private datasets [19] including nearly 2000 samples. Parashar et al. [20] proposes a two-stage method to classify glaucoma on the HRF [21] dataset. Corolla [13] employs self-supervised learning in their method, and all experiment is evaluated on the GAMMA [11] dataset.

Table.I shows the comparisons between ELF and other SOTA methods on glaucoma grading. The Corolla and our ELF are glaucoma grading for multi-modality, while others are evaluated on the single-modality dataset. Multi-modal can utilize the information between different modal data, such as fundus images and OCT volumes. Multi-modality information can significantly improve accuracy for glaucoma grading. Our proposed ELF is much better than other four compared methods, in term of both accuracy and Kappa. However, it is worth noting that the comparisons between different datasets are not fair enough. The multi-modal glaucoma grading is less researched since its novelty, so we can only compare with the Corolla method in the same dataset.

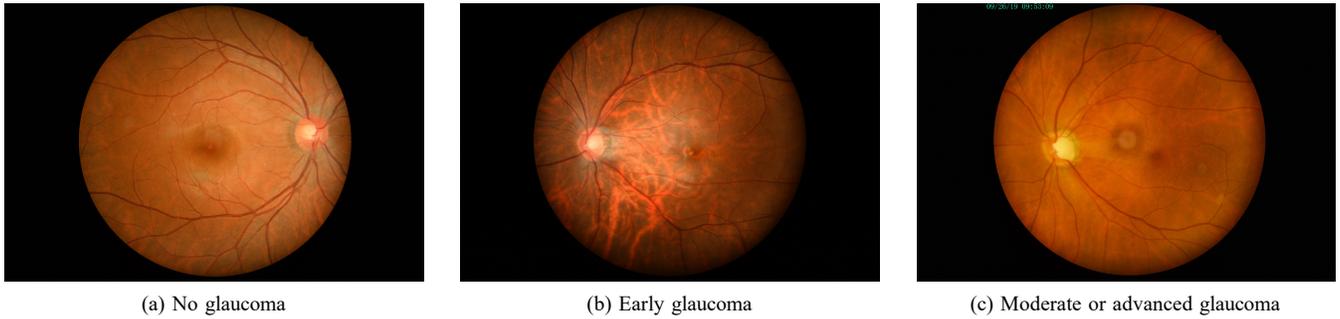

(a) No glaucoma      (b) Early glaucoma      (c) Moderate or advanced glaucoma

Fig. 2: Different stage of fundus photographs in the GAMMA dataset.

TABLE II: Ablation study of our proposed method on the glaucoma grading, in terms of Acc and Kappa. The highest score is shown in **boldface**.

| Fundus | OCT | LM | GM | Acc | Kappa |
|---|---|---|---|---|---|
| ✓ | | | | 0.790 | 0.673 |
| | ✓ | | | 0.740 | 0.575 |
| ✓ | ✓ | | | 0.810 | 0.702 |
| ✓ | ✓ | ✓ | | 0.900 | 0.853 |
| ✓ | ✓ | | ✓ | 0.890 | 0.819 |
| ✓ | ✓ | ✓ | ✓ | **0.930** | **0.896** |

*E. Ablation Study*

In Table.II, we present the ablation analysis results. We conduct the study of single-modality setting and the multi-modality setting seprately. First, we employ the ResNet and R(2+1)D network as the classifier with single-modality of color fundus photography and 3D OCT. Then, we use ResNet and R(2+1)D network simultaneously, the two features from two modalities are concatenated together to construct a joint classifier. To evaluate the effectiveness of our proposed method, we feed the local-wise feature in Eqn.6 the global-wise feature in Eqn.8 seprately to classifiers. Finally, we feed the classifier with concatenated local-wise and global-wise features to get the best Kappa score.

It is clear that when compared with single-modality methods, multi-modality methods can fully utilze the multi-modal information. A combination of fundus images and OCT volumes can acheive better performance than single-modal one. In addition, the ablation analysis can prove the effectiveness of our proposed method.

## IV. CONCLUSION

In this work, we proposed and evaluated an end-to-end local and global multi-modal fusion framework for glaucoma grading. In this framework, the local-wise and global-wise mutual information between fundus images and OCT volumes is extensively explored to improve the glaucoma grading accuracy. In addition, we conducted an extensive experiment on the only multi-modal GAMMA dataset. The result shows that our ELF method is much better than other SOTA methods. However, as a potential limitation of this work, the size of the multi-modal GAMMA dataset is relatively small. The limited size will lead to overfitting and poor robustness. Some techniques in the community of few-shot learning can alleviate this limitation of the GAMMA dataset. That would be the topic of our future work in multi-modal glaucoma grading for the small-size dataset.

ACKNOWLEDGMENT

This work is partly supported by the Science and Technology Development Fund, Macau SAR, under Grant 0087/2020/A2, MYRG2022-00190-FST.